# Study of the Backgrounds for the Search for Proton Decay to $10^{35}$Y at the WIPP Site with the LANNDD Detector

David B. Cline, Kevin Lee, Youngho Seo[*], and Peter F. Smith[1]

*Department of Physics and Astronomy, University of California, Los Angeles, California 90095-1547*

## Abstract

We briefly describe the LANNDD 70-kT liquid argon TPC proposal for the WIPP underground facility at Carlsbad, New Mexico. We, then, identify the key backgrounds for the search for $p \to K^+ \bar{\nu}$ to $10^{35}$ years lifetime. The most serious non-neutrino background is due to high-energy neutrons producing strange particles in the detector. We show that this can be reduced to an acceptable level by appropriate fiducial volume cuts.

## I    Current estimate of backgrounds

The 70-kT liquid argon LANNDD detector can be located at the underground science laboratory at the WIPP site as long as the depth is not a major concern for its scientific programs that include proton decay, solar and supernova neutrinos, atmospheric neutrinos. The detector is also able to function as an end detector with superbeam from the possible neutrino factories [1]. Of the potential physics objectives, the proton decay study to $10^{35}$ years is most sensitive to the site depth. Most of the background signals come from atmospheric neutrinos. According to recent ICARUS studies [2], we expect very low background levels for some of key proton decay channels (for example, $p \to e^+ \pi^0$, $p \to K^+ \bar{\nu}$, and $p \to \mu^- \pi^+ K^+$) up to $10^{35}$ nucleon years. Since the LANNDD at the WIPP facility most resembles the ICARUS T2400 at the Gran Sasso national laboratory in Italy [3], it is appropriate to look through the studies on background events that compete with proton decay signals. The following is a summarized discussion from the memorandum with emphasis on relevant aspects that are also applied to the LANNDD detector and also a smaller-version Mini-LANNDD detector [4].

*1.1. Nuclear and detector effects on signal events*

The ICARUS collaborators adopted standard GEANT-based FLUKA simulation codes with a nuclear interaction generator code called PEANUT. The two nuclear interactions, which are treated with special interests, inside the recoil nucleus are energy and momentum distributions due to Fermi motion and re-scattering of decay particles with nucleus before escaping the nucleus. These nuclear effects on potential nucleon decays distort two significant characteristics of a possible signal from a free nucleon, which are a fixed value of the total energy and the fact that the total momentum of the decay products is zero.

The effect due to Fermi motion results in spreading the spectrum of the invariant mass of bound nucleon decay products over a range about 40-MeV. A partial energy loss of decay products by re-scattering in the nuclear medium is the other important nuclear effect that cannot be ignored. In extreme cases, decay products are absorbed into the nucleus that is responsible their creation, which is crucial when the decay products include pions. However, when $K^+$ is contained among decay products, the nuclear interaction due to re-scattering is very weak, which corresponds to a high intrinsic survival fraction of signal.

**Figure 1** shows clear 40-MeV spreads of spectra due to Fermi motion only and absorptions of decay products with pions (significant) and kaons (weak). Kinetic variables (total measured energy, invariant mass, and total momentum) are, then, the bases of cuts to pick up proton decay signals from background events since they produce separable signatures for most-sought proton decay modes ($p \to e^+ \pi^0$ and $p \to K^+ \bar{\nu}$) when combined.

Detector effects on signals are included in the simulation codes, considering the detector geometry and the liquid argon response. Also, particle detection efficiency is assumed to be 100%, only requiring the acceptable minimum values of both momentum and kinetic energy.

---

[*] corresponding author e-mail: yseo@physics.ucla.edu
[1] also at *Rutherford Appleton Laboratory*, UK

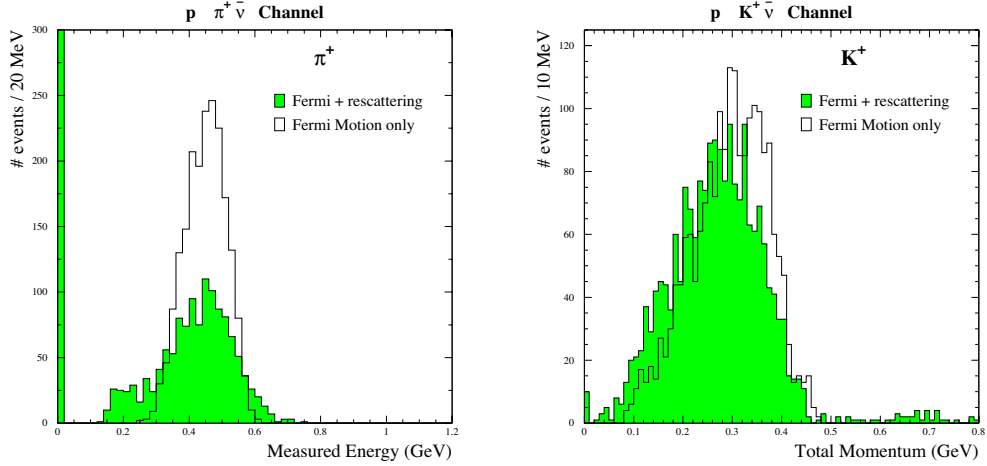

**Figure 1:** Nuclear effects on proton decay modes, $p \to \pi^+ \bar{\nu}$ (left) and $p \to K^+ \bar{\nu}$ (right). The filled histograms include both Fermi motion and re-scattering effects, and the open histograms are plotted for the cases with Fermi motion only. 40 MeV spread is clearly seen in open histograms for both proton decay channels. Figure is taken from [2].

*1.2. Atmospheric neutrino background estimation*

Atmospheric neutrino backgrounds are generated on a chosen primary nucleon by FLUKA codes. The number of generated neutrino-produced background events per each nucleon decay channel is amplified to the expected event rate for 1 MegaTon year exposure in order to obtain realistic estimates (in **Table 1**).

|  | $\nu_e$ CC | $\bar{\nu}_e$ CC | $\nu_\mu$ CC | $\bar{\nu}_\mu$ CC | $\nu$ NC | $\bar{\nu}$ NC |
|---|---|---|---|---|---|---|
| **# of Events** | 59861 | 11707 | 106884 | 27273 | 64705 | 29612 |

**Table 1**: Total number of simulated neutrino-produced background events per channel (equivalent to the expected event rate at 1 MegaTon year exposure). Numbers in the table are taken from [2].

In order to apply kinematic cuts, the total momentum, the total energy, and thus the invariant mass are computed for both signals and backgrounds, excluding protons, neutrons, and heavy remnants.

*1.3. Key proton decays with background estimates*

The proton decay selections are made with both topological cuts (search for exclusive final states) and kinematic cuts. Topological cuts may include both inclusive (decay products are absorbed) and exclusive (the presence of decay products is strictly required) final states, especially with pions as decay products. Hence, the intrinsic survival fractions (escaping probability out of the nucleus) are always noted when each cut is made.

For the $p \to e^+ \pi^0$ channels, the final state pions are absorbed in the nucleus with probability 45% (inclusive), while exclusive channel returns 54% intrinsic survival rate. Four topological cuts are applied for the exclusive channel: one pion, one positron, no charged pions, and no protons. Two kinematic cuts are: total momentum is less than 0.4 GeV, and total energy is between 0.86 GeV and 0.95 GeV. The inclusive decay channel exhibits dominant background events even after several topological and kinematic cuts.

**Table 2** lists background events with the specific cuts made. A simple interpretation from the numbers given in tables is that one background survives from 1 MegaTon year exposure after all six cuts, still keeping 45.30% of the signal efficiency for the exclusive $p \to e^+ \pi^0$ channel in **Table 2**.

| Exclusive channel cuts | $p \to e^+\pi^0$ | $\nu_e$ CC | $\bar{\nu}_e$ CC | $\nu_\mu$ CC | $\bar{\nu}_\mu$ CC | $\nu$ NC | $\bar{\nu}$ NC |
|---|---|---|---|---|---|---|---|
| One pion | 54.00% | 6604 | 2135 | 15259 | 5794 | 8095 | 3103 |
| One positron | 54.00% | 6572 | 2125 | 20 | 0 | 0 | 0 |
| No charged pions | 53.90% | 3605 | 847 | 5 | 0 | 0 | 0 |
| No protons | 50.85% | 1188 | 656 | 1 | 0 | 0 | 0 |
| Total momentum < 0.4 GeV | 46.70% | 454 | 127 | 0 | 0 | 0 | 0 |
| 0.86 GeV < Total energy < 0.95 GeV | 45.30% | 1 | 0 | 0 | 0 | 0 | 0 |

**Table 2:** Cuts for the exclusive $p \to e^+\pi^0$ channel. The first column indicates the survival fraction after each cut. The background events are listed correspondingly. The entire table was taken from [2].

Kinematic cuts are made using total momentum and invariant mass. **Figure 2** indicates that the selected region contains the accepted events inside the band (0.86 GeV < total energy < 0.95 GeV).

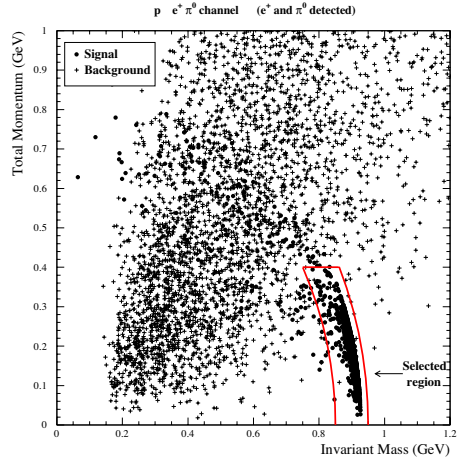

**Figure 2:** Kinematic cut in the exclusive $p \to e^+\pi^0$ channel. All events inside the band drawn (0.86 GeV < total energy < 0.95 GeV) are accepted. Figure is taken from [2].

For the $p \to K^+\bar{\nu}$ channel in which a strange meson (kaon) can be easily identified by the liquid argon TPC's excellent imaging capability and superb energy resolution, a very good survival efficiency (up to ~97%) can be obtained with only one topological cut (one kaon) and one kinematic cut (total energy < 0.8 GeV). This feature makes this proton decay mode be the most sought signal. **Table 3** lists the levels of background events with each cut made. **Figure 3** indicates that the kinematic cut (total energy < 0.8 GeV) is very efficient so that all events in the selected region can be accepted.

| Cuts | $p \to K^+\bar{\nu}$ | $\nu_e$ CC | $\bar{\nu}_e$ CC | $\nu_\mu$ CC | $\bar{\nu}_\mu$ CC | $\nu$ NC | $\bar{\nu}$ NC |
|---|---|---|---|---|---|---|---|
| One kaon | 96.75% | 308 | 36 | 871 | 146 | 282 | 77 |
| No pion | 96.75% | 143 | 14 | 404 | 56 | 138 | 25 |
| No positrons | 96.75% | 0 | 0 | 400 | 56 | 138 | 25 |
| No muons | 96.75% | 0 | 0 | 0 | 0 | 138 | 25 |
| No charged pions | 96.75% | 0 | 0 | 0 | 0 | 57 | 9 |
| Total energy < 0.8 GeV | 96.75% | 0 | 0 | 0 | 0 | 1 | 0 |

**Table 3:** Cuts for the exclusive $p \to K^+ \bar{\nu}$ channel. The first column indicates the survival fraction after each cut. The background events are listed correspondingly. The entire table was taken from [2].

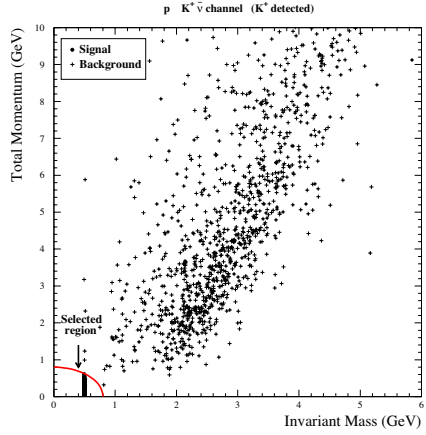

**Figure 3:** Kinematic cut in the $p \to K^+ \bar{\nu}$ channel. All events in the selected region (total energy < 0.8 GeV) are accepted. Figure is taken from [2].

*1.4. The summary table of the neutrino-produced backgrounds for the main proton decay search channels*

| Channel | Efficiency (%) | Background (5 kTon year) |
|---|---|---|
| $p \to e^+ \pi^0$ | 45.30 | 0.005 |
| $p \to e^+ (\pi^0)$ | 15.10 | 9.73 |
| $p \to K^+ \bar{\nu}$ | 96.75 | 0.005 |
| $p \to \mu^- \pi^+ K^+$ | 97.55 | 0.005 |
| $p \to e^+ \pi^+ \pi^-$ | 18.60 | 0.125 |
| $p \to e^+ \pi^+ (\pi^-)$ | 29.50 | 6.01 |
| $p \to e^+ (\pi^+ \pi^-)$ | 16.30 | 19.68 |
| $p \to \pi^+ \bar{\nu}$ | 41.82 | 3.91 |
| $p \to \mu^+ \pi^0$ | 44.80 | 0.04 |
| $p \to \mu^+ (\pi^0)$ | 17.85 | 20.81 |

Note that very high efficiencies coupled with very low background for some channels including kaons in the final state. It is possible to discover proton decay at the one-event level for these decay channels.

## II  New backgrounds in the $p \to K^+ \bar{\nu}$ channel from high-energy neutrons

High-energy neutrons could give a very serious, non-neutrino-induced, background to the proton decay ($p \to K^+ \bar{\nu}$) in the process:

$$n + p \to K^+ + [\Lambda, \Sigma] + neutral$$

with the Lambda decaying mainly into proton or neutron. High-energy neutrons can enter the detector with several quasi-elastic interactions and penetrate deeply. This background is clearly depth-dependent.

There are two processes to produce $K^+$ particles in nucleon-nucleon collision. One is a one-step direct collision of proton and nucleon to result in a $K^+$ and a hyperon (either $\Lambda$ or $\Sigma$), and a nucleon, which is described above. The other process, mostly responsible for subthreshold production of $K^+$, is a two-step cascade process that involves either/both $\Delta$ or/and $\pi$ in the intermediate state [5]. The cross sections for subthreshold production $K^+$ is well below $10^{-4}$ mb, but steeply approaches to zero at below 1-GeV energies for both experiments and calculations.

The measured total cross sections in $K^+$ production in $p + C$ collision, similar to $n + nucleon$ collision of our interest, are shown in **Figure 4**. The cross sections in $p+p$ and $p+n$ collisions can be assumed from the experimental agreement of the $p+C$ data and the scaled $p+p$ data ($\sigma(pp \to K^+) = \sigma(pn \to K^+)$)[5]. The cross sections for $p+n$ collision can be also scaled from the scale factor $\sigma_C^{inel} / \sigma_p^{inel} = 7.0$. We assume 0.1 mb for cross sections for $p+n$ collision to produce $K^+$, which is an order of magnitude lower from the value at 2.5-GeV energy in **Figure 4**. The cross section that we need in the collision $n+Ar$ for the LANNDD detector is also assumed to be at 1-mb level because this cross section must be a little bigger than that of $p+C$ collision.

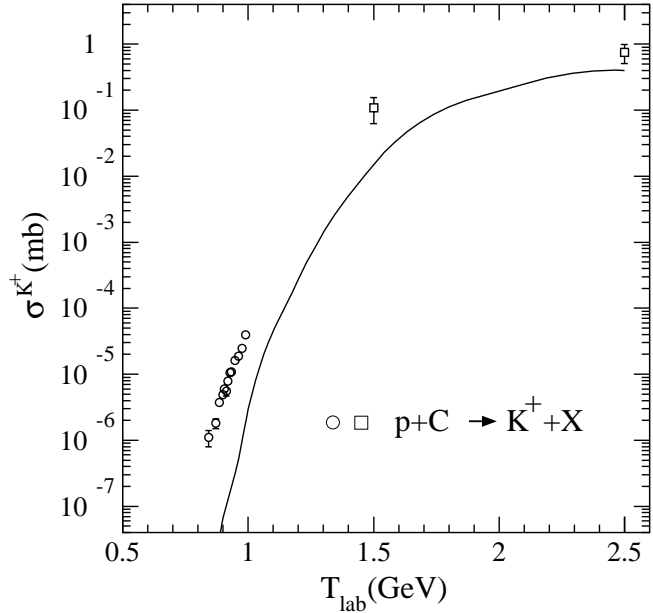

**Figure 4:** Total cross sections for $K^+$ production in $p+C$ collisions for the first chance collision compared to the experimental data. Figure is taken from [5].

Compared with atmospheric neutrino backgrounds, the background reaction has not very well studied yet. Clear understandings of the nuclear reaction of high-energy neutrons with nucleons in liquid argon (in the detector), and cross-sections as a function of energy, and the roles of both topological and kinematic cuts will further be investigated. The initial studies with neutron fluxes, dependent on depths, from a Monte Carlo simulation show that the depth-dependence is most crucial down to 2000 mwe (equivalent to the depth of the Carlsbad Underground Laboratory). A deeper site (e.g. 4000 mwe for comparison) reduces this background level, mainly due to the neutron flux reduction depending on depth, but not very significantly.

## III Depth-dependence of the $n + p \to K^+ + [\Lambda, \Sigma] + neutral$ background

First, the neutron fluxes are estimated from known production rates at depths 20 mwe, 2000 mwe, and 4000 mwe (**Figure 5**). Muons produce neutrons by absorption and spallation, and an in-house Monte Carlo FAUST was used to track these through the rock into the cavern. The points in **Figure 5** show the total per square cm per second in each factor 10 energy span. Note that there is a large difference between neutron fluxes at 20 mwe and 2000 mwe, but less than an order of magnitude between 2000 mwe and 4000 mwe.

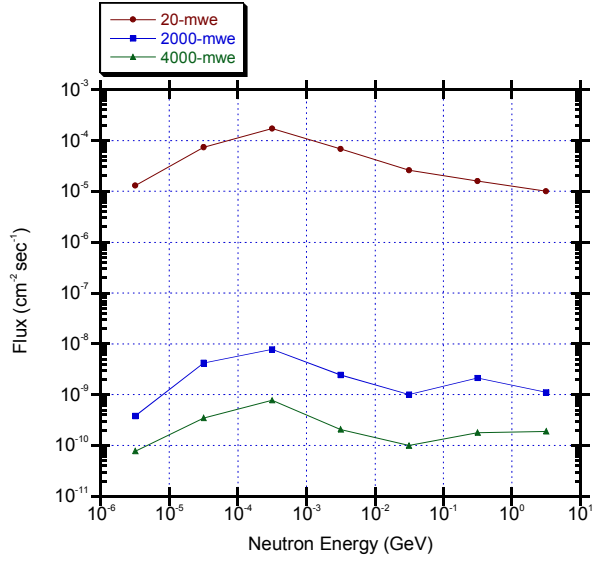

**Figure 5**: Estimated muon-related neutron fluxes for depths 20-mwe, 2000-mwe, and 4000-mwe. There is an additional site-dependent neutron flux from U and Th in the rock, but this is below 10 MeV and does not contribute to this background.

From the estimated fluxes, the rate is calculated as flux × cross section (per atom) × $6\times10^{23}$/A × mass, with A=40 for Ar. The cross section is taken as 1 mb and the effective mass 19 kT (out of 70 kT Ar mass for the LANNNDD detector), for a typical 1-m mean neutron penetration into the argon. The calculated backgrounds from the $n+\text{Ar} \rightarrow K^+ +[\Lambda,\Sigma]+ neutral$ process are shown in **Table 4**. Note that the 1-year backgrounds for this outer 19 kT of the volume are comparable to the backgrounds due to atmospheric neutrinos in **Table 1** (for 1 MegaTon year exposure) for GeV-range neutrons – i.e. those above the threshold for this background.

The calculated backgrounds from $n+p \rightarrow K^+ +[\Lambda,\Sigma]+ neutral$ process are in **Table 4**. Note that the backgrounds of the reaction for 19 kT year exposure are comparable to the backgrounds due to atmospheric neutrinos in **Table 1** (for 1 MegaTon year exposure) for GeV-range neutrons, which are of fundamental significance for this background since the threshold to produce $K^+$ in the free nucleon-nucleon collision is about 1.5 GeV.

| **Neutron energy** | **20-mwe** | **2000-mwe** | **4000-mwe** |
|---|---|---|---|
| 1 to 10 GeV | $5\times10^5$ | 5000 | 1000 |

**Table 4**: Calculated background events without any cuts for the process $n+p \rightarrow K^+ +[\Lambda,\Sigma]+ neutral$ with 1-mb cross section. The energy threshold for the free nucleon-nucleon collision to produce $K^+$ responsible for this background reaction is about 1.5 GeV. The background can be reduced by as much as a factor of 1000 with both topological and kinematic cuts.

There are some other factors that could reduce the level of backgrounds:

1) These backgrounds will be sensitive to topological cuts that require one $K^+$ and no other final state particles because other particles (that can escape the detector undetected, though) are produced at the vertex. Also, the kinematic cuts must be very effective to filter out these backgrounds since the $K^+$ will share the beam momentum, making the momentum carried by $K^+$ will be large, so that the fraction fitting the kinematics for proton decay events will be reduced as the neutron energy increases.

2) The assumption that this background occupies the outer 19 kT is an over-estimate, since penetration of the neutrons into the volume is accompanies by multiple scattering and energy loss to below the 1.5-GeV threshold. More detailed Monte Carlo studies will give the effective penetration for which the neutron energy remains > 1.5 GeV.

3) Eliminating this outer region from the fiducial volume will thus reduce the background within the remaining fiducial volume by many orders of magnitude.
4) For initial studies of the LANNDD detector, a smaller module, the 5-kT Mini-LANNDD, will be first tested. For the smaller-sized detector, the effective volume will be substantially low, compared to 19 kT from the 70-kT LANNDD. The effective volume (1-m cut from the wall) for the 5-kT Mini-LANNDD is approximately 1.5 kT.

The above factors will reduce the background by a large factor, to the extent that the difference between backgrounds at 2000 mwe and 4000 mwe is not significant, and there is no need to operate the LANNDDD detector below the Carlsbad depth. This is also true for the neutrino background that is independent of depth. However, the background at the surface (20 mwe) would make it impossible to search for proton decay at this level, which in addition could not be distinguished from the large flux of cosmic ray events.

**Acknowledgements:** We thank Andre Rubbia, Franco Sergiampietri, and the members of ICARUS team for their help. Professor Peter F. Smith acknowledges partial support from an Individual Emeritus Research Award by The Leverhulme Trust.

# References

[1] D. B. Cline *et al.*, "Proposal to study the feasibility to site various neutrino detectors at WIPP for neutrino factories or superbeams", *proposal to DOE*, 2002.

[2] A. Bueno, A. Ferrari, S. Navas, A. Rubbia, and P. Sala, "Nucleon decay searches: Study of nuclear effects and backgrounds", *ICARUS/TM-2001/04*, 2001

[3] D. B. Cline, B. Lisowski, C. Matthey, S. Otwinowski, Y. Seo, F. Sergiampietri, X. Yang, and H. Wang, "Proposal to construct the HV system for ICARUS T2400 at LNGS", *proposal to DOE*, 2002.

[4] D. B. Cline, S. Otwinowski, and F. Sergiampietri, "Mini-LANNDD: A very sensitive neutrino detector to measure $sin^2(2\theta_{13})$, *astro-ph/0206124*.

[5] M. Debowski et al., "Subthreshold K+ production in proton-nucleus collisions", *Z. Phys. A 356*, 313-325, 1996.